# Designing the Hybrid Cooperative: A Socio-Technical Architecture for Scalable, Global Coordination Using Blockchain

Henrik Axelsen
Copenhagen Business School
ha.digi@cbs.dk

Jan Damsgaard
Copenhagen Businss School
jd.dig@cbs.dk

**Abstract**

*Blockchain has been promoted as a remedy for coordination in fragmented, multi-stakeholder ecosystems, yet many projects stall at pilot stage. Using a design-science approach, we develop the Hybrid Cooperative (HC), a digitally native governance architecture that combines smart-contract coordination with a minimal, code-deferent legal interface and jurisdictional modules. This selective decentralization decentralizes rules where programmability lowers agency and verification costs, and centralizes only what is needed for enforceability. A post-case evaluation against two traceability initiatives in supply chains illustrates how the HC improves distributed task management, verifiable information, incentive alignment, institutional interoperability, and scalable, contestable governance. The paper contributes to Information Systems by specifying a socio-technical model for scalable, multi-stakeholder coordination across regulatory and organizational boundaries.*

## 1. Introduction

Supply chains are neither classic hierarchies nor fully fledged platforms, but hybrid networks of semi-autonomous firms coordinated more through negotiation than control (Powell, 1990).

Despite blockchain's promise of transparency and traceability, most initiatives in supply chain coordination have failed to move beyond pilots (Sternberg et al., 2021). The impediment is less about technical maturity than about coordinated action across fragmented legal, institutional, and operational contexts. In shipping and gemstones, for example, technically robust systems faltered due to governance asymmetries, incentive misalignment, and weak institutional anchoring (Prockl et al., 2022). Addressing incentive structures for collaboration at scale is therefore a core scientific and practical challenge, with implications for how large digital infrastructures are organized.

Two digital governance forms have emerged as alternatives to firm-centric coordination. Decentralized Autonomous Organizations (DAOs) promise on-chain decision making through smart contracts and token voting (Ellinger et al., 2024; Lumineau et al., 2021). Platform cooperatives extend democratic ownership and community governance to digital contexts (Zhu & Marjanovic, 2020, 2024). Yet neither model, in its current instantiation, resolves the coordination demands of complex, transnational ecosystems. DAOs frequently lack legal enforceability and mechanisms for exception handling; they presuppose that organizing logic can be fully codified, which clashes with the institutional embeddedness and interpretive flexibility observed in real organizations (Axelsen et al., 2024; He & Puranam, 2023; Orlikowski, 1992; UK Law Commission, 2024). Platform cooperatives, while institutionally legitimate, often remain jurisdiction-bound and technologically less composable at scale (ICA, 2015; Zhu & Marjanovic, 2024).

In this landscape cooperative governance offers a durable, member-oriented alternative to investor-owned firms (Fairbairn, 1995), designed around member benefit and fiduciary duties to users, not external capital, supporting long time horizons, reciprocal monitoring, and resilience across cycles (ICA, 2015). In cross-firm settings, cooperative principles – democratic control, proportional returns, and shared purpose – map naturally to the core coordination problems of task division, information provision, and reward distribution.

The cooperative movement's longevity and accountability provide institutional grounding; our aim is to retain these virtues while enabling programmable, cross-jurisdictional coordination. What is then needed is a digital, socio-technical, governance model that preserves cooperative legitimacy while adding programmability and modular legal interfaces for transnational operation. This socio-technical governance architecture should be tailored to decentralized supply chains, that is, cross-firm networks without a single apex authority but with persistent interdependence. By socio-technical, we mean the deliberate integration of social structures – participants, roles, institutions, and legal forms – with technical systems – smart contracts, consensus, and tokens – as co-evolving components of governance (Orlikowski, 1992).

Our research question is therefore: *How can a digital governance model enable trustworthy, inclusive coordination among semi-autonomous actors in complex supply chain ecosystems?*

To address this question, we develop a conceptual design artifact that formalizes coordination across three interlocking layers: a programmable layer for token-based coordination, that is, the use of programmable tokens as rights and incentives to allocate access, tasks, and rewards; a legal foundation that supplies enforceability and regulatory interface; and jurisdiction-specific modules that adapt participation to local legal requirements. Aligning participation at scale requires tokenized incentive systems, meaning rules that issue, vest, and redeem tokens to align contributions with benefits, and scalable consensus, that is, agreement on shared state and decisions that remains reliable as participation grows. In combination, these layers target the persistent bottlenecks reported in prior deployments – concentrated control, missing shared rules, and institutional fragmentation – while preserving stakeholder autonomy and legal enforceability.

Our contribution is twofold. First, we synthesize insights from information systems, organizational theory, and blockchain governance to propose the Hybrid Cooperative (HC) as a conceptual design artifact. Second, we conduct a post-case evaluation using two relevant case studies – TradeLens and Provenance Proof (Prockl et al., 2022), illustrating how the artifact addresses failure modes documented in these deployments. By articulating a new organizing form for digitally enabled supply chains, we add to ongoing Information Systems (IS) debates on socio-technical governance, hybrid organizing, and the institutional evolution of digital infrastructures. The HC advances the conversation on how programmable architectures can support resilient, transparent ecosystems in contexts that demand both automation and institutional legitimacy.

The remainder of the paper proceeds as follows: Section 2 reviews related work; Section 3 derives the design requirements; Section 4 specifies the HC architecture; Section 5 presents the evaluation; Section 6 discusses implications; and Section 7 concludes.

## 2. Related work

To ground our design work, this section reviews organizing forms and governance mechanisms for digital coordination in decentralized supply chains, that is, cross-firm networks with no single controlling authority but sustained interdependence across legal and institutional boundaries – and not necessarily completely decentralized or autonomous, but sufficiently decentralized to avoid central control by dominant actors. We synthesize insights on (i) governance and coordination problems, (ii) blockchain governance and smart contracts, (iii) organizational forms for digital coordination, (iv) DAO potential and limits, (v) platform cooperatives and hybrid organizing, and (vi) legal wrappers and institutional interoperability, and conclude with the design gaps our Hybrid Cooperative (HC) addresses.

(i) Decentralized supply chains resemble hybrid networks of semi-autonomous firms coordinated more through negotiation than hierarchy (Powell, 1990). Organizational theory holds that any viable arrangement must solve task division and allocation, information provision, and reward distribution (Puranam et al., 2014; Williamson, 1981). Empirically, blockchain pilots stall less for technical reasons than because these coordination problems collide with institutional fragmentation, asymmetric control, and misaligned incentives (Prockl et al., 2022; Sternberg et al., 2021). The design challenge is therefore socio-technical: align rules, roles, and incentives across organizations and jurisdictions while maintaining legal enforceability and adaptability. Platform-led supply-chain initiatives such as TradeLens illustrate how orchestrator control and governance opacity can inhibit third-party participation even with technically mature infrastructure (Jovanovic et al., 2022; Prockl et al., 2022).

(ii) Blockchains enable coordination without a centralized authority by encoding rules as executable smart contracts, providing tamper-evident, auditable transaction histories, and making rights programmable via tokens (Beck & Müller-Bloch, 2018). Smart contracts automate core coordination tasks, including access and role management, workflow triggers, dispute-handling hooks, and token-based coordination, that is, using programmable tokens as rights and incentives to allocate access, tasks, and rewards (Gregory et al., 2024; Lumineau et al., 2021). The promise of scalable consensus – agreement on shared state and decisions that remains reliable as participation grows – suggests potential reductions in agency and verification costs (Davidson, 2023). Yet automation alone cannot substitute for interpretation, exception handling, and institutional recourse in cross-jurisdictional contexts (He & Puranam, 2023). Studies in supply chain management document interest in these mechanisms, alongside limits to durable, inclusive governance at scale (Beck et al., 2023; Beck & Müller-Bloch, 2017; Pflaum et al., 2024; Schwiderowski et al., 2024; Swain & Patra, 2022).

(iii) Traditional firms provide legal personhood, finance, and operational discipline but hinder broad participation and cross-jurisdictional adaptability (Ixmeier et al., 2024). Nonprofits and foundations offer mission alignment and public accountability yet face rigid charters and weak mechanisms for participatory governance or tokenized incentives (Ebrahim, 2003; Guo & Acar, 2005). Open-source communities excel at collaborative innovation and knowledge sharing but typically lack legal status, sustainable financing, and accountability structures needed for complex, capital-intensive public goods (O'Mahony & Ferraro, 2007; World Bank Group, 2019). These forms illuminate trade-offs among participation, enforceability, and scalability that any digital governance model must reconcile.

(iv) DAOs shift decision rights on-chain via token voting and modular processes (Ellinger et al., 2024; Lumineau et al., 2021) but face voter apathy, token-holder capture, and legal ambiguity that complicates contracting, liability, and regulatory interface (Axelsen et al., 2024; UK Law Commission, 2024; Wright, 2021). High-profile experiments such as KlimaDAO show how token fragility and unclear legal standing can derail missions even when on-chain mechanics work (Jirásek, 2023). In short, DAOs offer strong automation but weak enforceability and limited off-chain integration for transnational supply chains, echoing IS insights that organizing cannot be fully codified and remains institutionally embedded (Orlikowski, 1992).

(v) Platform cooperatives aim to translate cooperative principles – democratic control and member ownership – into digital contexts (ICA, 2015; Zhu & Marjanovic, 2024). They provide institutional legitimacy and a clear social purpose but often rely on centralized infrastructure and jurisdiction-bound legal forms, which constrains cross-border scalability and composability (Battilana et al., 2015; Carballo, 2023; Ladd et al., 2024; Santana & Albareda, 2022; Shepherd et al., 2019).

(vi) Legal wrappers, that is, lightweight entities enclosing treasury, IP, or contracting, give code-governed communities legal personhood to interface with external institutions (Brummer & Seira, 2022). (UK Law Commission, 2024) maps a spectrum of wrapped/adjacent configurations and highlights code deference, that is, governing documents that commit the legal entity to implement on-chain outcomes within applicable law. These approaches trade some decentralization for enforceability and regulatory interface—the very trade-offs decentralized supply chains confront when scaling beyond pilots.

Across these six streams, three observations recur. First, blockchain's value here lies less in infrastructure per se than in governance mechanisms that make coordination credible and auditable. Second, neither DAOs nor platform cooperatives alone resolves the tension between programmability and institutional legitimacy at transnational scale. Third, legal wrappers provide enforceability but risk recentralization without careful interfaces to on-chain governance. These observations motivate a hybrid socio-technical architecture integrating smart-contract coordination, legal agency, and jurisdictional modularity – the design space our Hybrid Cooperative (HC) aims to operationalize. We adopt a selective decentralization stance: the goal is not maximum decentralization but credible, enforceable decentralization, that is, decentralize the parts of governance where programmability reduces agency and verification costs while retaining minimal, purpose-bound legal interfaces where contractual capacity and accountability are indispensable. The trade-off is explicit: some centralization is introduced to obtain legal agency; in exchange, governance logic remains auditable, rule-bound, and contestable on-chain rather than opaque and discretionary off-chain (Axelsen et al., 2024; Orlikowski, 1992; Wright, 2021). This stance frames the HC as a socio-technical architecture that decentralizes task and incentive coordination via smart contracts while centralizing only what is necessary for enforceability and multi-jurisdictional compliance through neutral, purpose-bound legal modules.

We next derive design criteria that reflect core coordination challenges and the practical requirements of scalable, legally grounded governance.

## 3. Design criteria

We adopt a Design Science Research approach (Gregor & Hevner, 2013; Hevner et al., 2004) to formulate requirements for a governance model that supports coordination in decentralized supply chains as defined earlier. To avoid arbitrary selection, we derive requirements through triangulation across three bodies of knowledge: (i) Organizing theory — any viable organizing form must solve task division and allocation, information provision, and reward distribution (Puranam et al., 2014; Williamson, 1981); (ii) Blockchain/DAO governance – programmable coordination can reduce agency/verification costs, but faces limits in codifiability, consensus scalability, and exception handling (Davidson, 2023; Ellinger et al., 2024; Gregory et al., 2024; He & Puranam, 2023; Lumineau et al., 2021); Legal-institutional design – cross-jurisdictional operation requires legal personhood, liability shielding, tax/regulatory fit, and code–law interfaces (Brummer & Seira, 2022; UK Law Commission, 2024; Wright, 2021).

We map the three coordination problems – task division and allocation, information provision, and reward distribution – to blockchain governance mechanisms (smart contracts, tokens, consensus) and to the legal capacities required for enforceability (legal personhood, liability shielding, code–law interfaces). This yields five mutually constraining requirements: use programmability where it reduces agency and verification costs and provides a legal interface where contractual capacity is indispensable. Table 1 consolidates these five interdependent requirements (R1–R5).

| Requirement | Requirement statement |
|---|---|
| R1: Distributed Task Management | Enable modular workstreams and dynamic role assignments; support non-hierarchical task division, delegation, and escalation among distributed stakeholders. |
| R2: Information Provision | Ensure near–real-time access to governance and operational data with on-chain visibility (e.g., transactions, votes) and verifiable off-chain attestations (e.g., sustainability, compliance). |
| R3: Aligned Incentive Structures | Provide programmable, multi-modal rewards that align stakeholder incentives – balancing token-based logic with cooperative values and rights-sharing. |
| R4: Institutional Interoperability | Maintain legal personhood and jurisdictional modularity to own assets, contract, satisfy compliance obligations, and minimize tax/regulatory friction. |
| R5: Scalable Governance Architecture | Remain operable in fragmented, uncertain environments, supporting coordination among semi-autonomous actors without relying on a centralized platform or firm control. |

Table 1. Design Criteria

Theoretically, R1–R3 respond to the canonical coordination problems – tasks, information, rewards – while R4–R5 add the legal and procedural scaffolding for enforceable, multi-party operation across jurisdictions. Practically, the failure modes observed in our two case studies as further detailed in the evaluation section below – weak incentive alignment, missing legal agency, and concentrated control – map to R3, R4, and R5 respectively.

The requirements are interdependent (e.g., stronger modularization in R1 raises demands on R2 interfaces and R5 load-balancing).

In the next section we instantiate R1–R5 in the Hybrid Cooperative's three layers (programmable, legal foundation with code-deference, jurisdictional modules) and make the trade-offs explicit.

## 4. The HC as a Socio-Technical Artifact

Building on the spectrum outlined earlier, we adopt a specialized form of the partially wrapped model discussed by (UK Law Commission, 2024), which identifies four structural models for DAO-related legal arrangements (or more broadly, smart contract-native organizations): (i) fully wrapped DAOs, where a legal entity (such as a company or foundation) is directly governed by token holders; (ii) partially wrapped DAOs, where a separate legal entity performs specific off-chain functions while remaining distinct from on-chain governance; (iii) DAO-adjacent entities, where traditional legal structures operate in parallel with DAO activity; and (iv) digital legal entities (e.g., companies, cooperatives, LLCs) that adopt smart contracts internally but retain off-chain governance. Platform cooperatives fall into this last category: they decentralize governance via democratic control and member ownership, yet remain bounded by national statutes, limiting technological and jurisdictional scalability.

In contrast, we adopt a specialized partially wrapped model centered on a Foundation Company. This structure preserves DAO-native governance for strategic decision-making while assigning the foundation responsibility for off-chain assets, compliance, IP, and contracts. Unlike typical partially wrapped DAOs, token-holder governance is embedded in the foundation's bylaws (code-deference), aligning decentralized coordination with legal authority. The model supports jurisdictional modularity, allowing local legal entities to be added as needed to comply with onshore requirements (e.g., EU Digital Product Passport, MiCA).

The HC adopts selective decentralization: governance logic remains programmable and auditable on-chain, while a minimal legal interface provides enforceable contractual agency. This justifies blockchain over a trusted database by enabling shared control of rules and treasuries, tamper-evident governance history, and programmable incentives under heterogeneous trust.

Although preferred jurisdictions shift over time, foundation-style wrappers (e.g., a Cayman Foundation Company under the Foundation Companies Act 2017 as referenced by (UK Law Commission, 2024)) are one option where by-laws can defer to code and provide legal personhood. Equivalent wrappers in other jurisdictions (e.g., foundation/association/cooperative forms) can be used without altering the HC's governance logic, provided they can commit to code-deference and a purpose-bound remit. Like other foundations in many jurisdictions it provides legal personhood, but it also provides a shareholder-free, purpose-driven form with bylaw customization – including token-holder voting – and board-based administration compatible with code-deference. It offers liability shielding and tax neutrality within a crypto-aligned regulatory regime, while enabling DAOs to shape the foundation's mission directly.

We implement this structure as the Hybrid Cooperative (HC) – a digitally native organizing form composed of three socio-technical layers that correspond to the design requirements (see Figure 1).

Layer 1 consists of optional jurisdictional modules (e.g., cooperatives, Special Purpose Vehicles (SPVs), subsidiaries) tailored to local regulatory or sector needs; this modularity enables onshore operations without redesigning the core. Layer 2 is the programmable governance layer, where smart contracts implement collective decision-making, task allocation, and incentive logic through token-based mechanisms; voting, treasury, and registry contracts provide transparent, auditable coordination, enacted via a DAO. Layer 3 is the legal foundation that provides personhood, manages off-chain assets, and ensures compliance capacity; its charter embeds code-deference, that is, a commitment to implement on-chain decisions within applicable law, with a purpose-bound remit (custody, IP/licensing, contracting, compliance) and no discretionary policymaking beyond on-chain authorization. The foundation has no shareholders; participants therefore obtain a stake not as equity in the foundation but via the HC governance token and the on-chain membership registry.

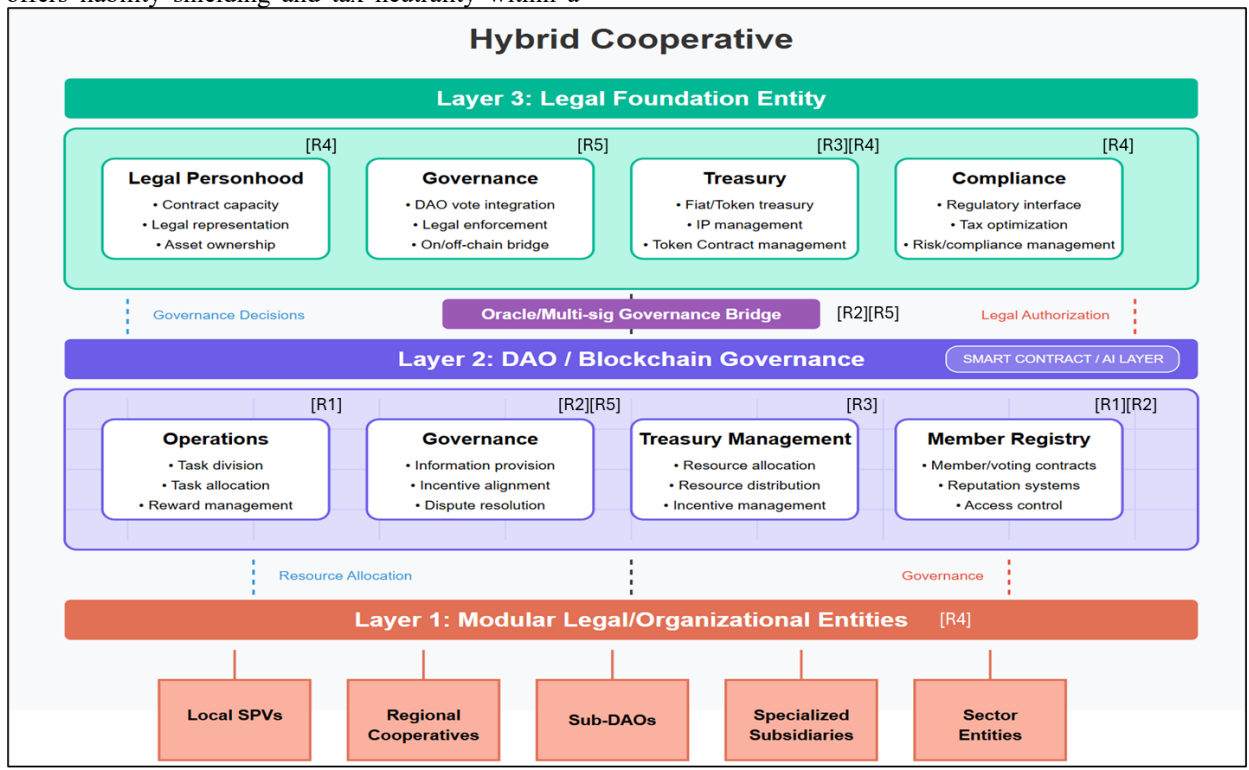

**Figure 1. The Hybrid Cooperative: A three-layered governance model and core contribution of this paper.**

Token distribution follows a published policy – initial allocation to founding contributors; ongoing issuance for verifiable contributions and programmatic grants; vesting/lockups and for-cause clawbacks to align long-term participation. On-chain voting power is used to elect/ratify/remove directors, approve treasury actions, and admit/exit jurisdictional modules, and directors are obliged by code-deferent bylaws to implement approved resolutions within applicable law.

Figure 1 legends trace back to the requirements as follows: R1 → workflow/role contracts; R2 → on-chain records + verifiable off-chain attestations; R3 → token issuance/vesting/redemption rules; R4 → foundation with code-deference + modular local entities; R5 → delegation/committees, quorum / supermajorities, upgrade & dispute procedures.

Smart contracts serve as a programmable governance substrate for task allocation, incentive distribution, voting, and dispute handling, reducing manual oversight and complex bilateral contracting. Crucially, they are not isolated from legal authority: under code-deference, legal wrappers (e.g., the foundation or DAO-controlled SPVs) must execute DAO-approved outcomes to the extent permitted by law. Trusted oracles transmit off-chain events (regulatory approvals, real-world transactions, dispute outcomes) on-chain; we use multi-provider oracles with aggregation rules and operator rotation. Changes to oracle sets/weights are governed on-chain; exceptional overrides require elevated thresholds and leave an auditable trace. Thus, programmable coordination is anchored in enforceable legal structures, making the HC suitable for policy-aligned and jurisdictionally complex operations.

By design, the HC embeds social logics – stakeholder participation, fiduciary accountability, community legitimacy – into technical governance. Stakeholders obtain rights via transparent onboarding defined in governance contracts. Directors are elected or ratified on-chain, with removal for breach of code-deference or fiduciary duty.

Major actions (e.g., treasury movements, parameter changes, module admission/exit) require on-chain approval with quorum/supermajorities to prevent capture. In practice, the HC enables token-based consumer and supply-chain engagement while providing trusted structures for IP management and regulatory interface across domains (retail, tourism, food, circular economy). Unlike platform-led systems, the HC separates policy from execution, removes privileged operator control, and makes authority procedurally contestable and auditable.

These choices target failure modes observed in prior deployments (incentive misalignment, missing legal agency, concentrated control) and are assessed next in the post-case evaluation.

Aligned with IS traditions of studying infrastructure (Henfridsson & Bygstad, 2013), we position the HC as a minimal viable institutional infrastructure for cross-sector coordination. With its modular, extensible architecture—grounded in technical automation and institutional legitimacy—the HC appears suited to contexts where existing forms fail to scale beyond pilots, such as decentralized supply chains and multi-stakeholder networks (Sternberg et al., 2021).

## 5. Evaluation

Table 2 maps the conceptual HC artifact design elements to a corresponding structural or governance element in the HC (DAO, foundation, legal modules).

| Requirement | Evaluation Summary |
| --- | --- |
| R1: Distributed Task Management | The DAO layer enables modular governance of functions such as token issuance, protocol maintenance, sustainability tracking, and grant allocation. Smart contracts support dynamic task allocation via token-weighted or role-based governance, while legal modules (e.g., SPVs or cooperatives) can hold specific responsibilities in parallel, minimizing coordination overhead. |
| R2: Information Provision | On-chain governance, treasury activity, and member registries provide transparent, auditable records of decisions and resource flows. Off-chain information (e.g., contracts, sustainability attestations, compliance reports) is handled by the foundation, enabling verifiable documentation and sharing. |
| R3: Reward Distribution | Incentives are programmable through smart contract mechanisms (e.g., staking, grants, airdrops), with legal entities able to distribute fiat or crypto-based rewards. This allows both mission-driven and token-based rewards to |

| Requirement | Evaluation Summary |
|---|---|
| | align diverse stakeholder incentives. |
| R4: Institutional Interoperability | A Foundation Company provides legal personhood, asset ownership, and contracting capacity on behalf of the DAO. Jurisdictional modularity is ensured through the ability to add local legal entities (e.g., EU-based cooperatives, traditional firms or SPVs) as needed, while tax neutrality supports cross-border engagement. |
| R5: Scalable Governance Architecture | The HC's three-layer structure (DAO logic, legal foundation, modular entities) supports coordination among semi-autonomous actors across fragmented environments. It enables governance across sectors and jurisdictions while adapting to regulatory frameworks like MiCA or the EU Digital Product Passport. |

**Table 2. HC evaluation against requirements.**

While the HC may enable democratic governance through token-weighted voting, role-based participation, and transparent smart contract design, its anchoring in a legal wrapper necessarily tempers the ideal of full decentralization and autonomy compared to the DAO ethos. The foundation's role as a legal interface introduces a degree of centralization that, while necessary for regulatory and operational engagement, also constrains pure on-chain governance, highlighting the inherent trade-off between enforceability and decentralization.

Consistent with FEDS (Venable et al., 2016), we conduct an ex post, naturalistic post-case evaluation of the conceptual HC design using two real deployments as episodes: TradeLens (TL) and Provenance Proof (PP) (Prockl et al., 2022); for TL also (Jovanovic et al., 2022).

We address three evaluation questions: (1) problem–requirement fit: Do the barriers observed in TL/PP correspond to R1–R5? (2) mechanism adequacy: Do the HC mechanisms plausibly mitigate those barriers? (3) differentiation: How does HC governance differ from TL/PP governance?

For procedure we conducted a structured reading of the case papers (and official materials for factual setup) to (i) extract issues; (ii) code each issue to one or more requirements R1–R5; and (iii) map to the corresponding HC mechanism(s), recording coverage as *Direct/Partial/Gap*. This naturalistic post-case strategy matches evaluation form to artifact maturity and assesses theory–practice fit without a new field instantiation.

Provenance Proof (PP) aimed to enhance traceability in the fragmented gemstone supply chain using digital twins. Despite technical viability, adoption lagged due to governance imbalances, opaque control by the initiator Gübelin Gem, and concerns over proprietary technology. Smaller actors hesitated to join, fearing dependency and lack of influence. TradeLens (TL), a joint venture by Maersk and IBM for container shipping transparency, faced similar adoption barriers. Although technically robust and widely onboarded, competitors resisted due to Maersk's dominant role and perceived control. Integration costs and governance opacity limited network effects and stakeholder trust.

Coverage judgments vs HC features:

R1: *Direct* (TL, PP). Where TL acted as a central orchestrator, PP operated with initiator-led permissions. In contrast, HC's programmable workflows, role/permission registries, delegation and escalation can mitigate the issue as stated.

R2: *Partial* (TL, PP). Where TL centralized data governance and PP used proprietary/limited visibility, HC's on-chain audit trails and verifiable off-chain attestations address opacity, but effectiveness depends on oracle network design and participant adoption of attestation processes.

R3: *Partial* (TL, PP). TL imposed high onboarding costs with unclear value share and PP concentrated benefits/control with the initiator. In contrast, HC supplies token issuance/vesting/redemption and optional cooperative dividends/IP-sharing, but incentive magnitudes, vesting horizons, and eligibility rules require context-specific calibration and legal constraints.

R4: *Direct* (TL, PP). TL's JV wrapper arrived late with perceived initiator advantage and PP lacked a neutral legal interface. HC's code-deferent foundation and jurisdictional modules directly address TL's late/advantaged wrapper and PP's missing neutral legal interface (contracting, compliance, IP).

R5: *Partial* (TL, PP). TL/PP governance remained static and initiator-led with limited contestability. HC introduces quorum/supermajorities, delegated committees, upgrade/dispute procedures, and anti-capture features; their efficacy depends on parameterization (thresholds, delegation scope) and participation levels.

Despite operating in different industries, both cases illustrate remarkably similar governance and structural barriers. In each, the presence of a dominant initiating actor raised concerns about control, lock-in, and long-term dependency. This imbalance eroded trust and discouraged broader adoption, particularly among smaller or competing actors. The absence of scalable governance frameworks and unclear mechanisms for value distribution and value capture limited both platforms' ability to scale coordination and sustain equitable stakeholder participation. While technically functional, they fell short due to governance misalignment and a lack of participatory structures capable of fostering trust and coordination at scale. These dynamics exemplify the "tragedy of the facilitated commons," where externally orchestrated collaborations, despite technical soundness, often fail to generate the autonomy and shared norms required for sustainable adoption (Sternberg et al., 2021). These shortcomings underscore the need for governance architectures that are better suited to decentralized, cross-organizational coordination.

Taken together, the HC differs by addressing the structural limitations identified in both case studies by offering a flexible yet institutionally grounded governance framework. While not a guaranteed solution, it represents a promising approach for enabling sustained coordination across semi-autonomous actors in digitally mediated supply networks.

## 6. Discussion

The TradeLens and Provenance Proof cases show how well-funded, technically sound initiatives can stall when governance is misaligned with institutional and competitive realities. Dominant initiators raised concerns about control and dependency, undermining trust and uptake. These challenges highlight the limits of technical decentralization when not matched with institutional legitimacy and participatory scaffolding.

In this research we adopt selective decentralization, that is, decentralize governance logic where programmability reduces agency and verification costs, and retain a minimal legal interface where contractual capacity and accountability are indispensable. The trade-off is explicit – some centralization is introduced to obtain legal agency – mitigated by code deference, a purpose-bound remit with no discretionary policymaking, quorum and delegation limits, and multi-provider oracle governance with auditable overrides.

Alternative models surface complementary strengths but also gaps at scale: Platform cooperatives provide democratic legitimacy yet are jurisdiction-bound and infrastructure-centric, limiting cross-border composability and access to capital. DAOs offer automation and transparency yet typically lack legal personhood for contracting, IP, and compliance. Both struggle to balance autonomy with enforceability and inclusivity with operational coherence – tensions that proved decisive in the gemstone and shipping platforms. In contrast, the Hybrid Cooperative (HC) combines programmable, role-based governance with a neutral legal foundation and optional local modules, addressing these tensions as a cohesive socio-technical design.

Adoption facilitators include a neutral, code-deferent wrapper; transparent value sharing via tokenized rewards and optional cooperative-style benefits; verifiable information flows (on-chain audit plus off-chain attestations); modular local entities for onshore contracting and compliance; and clear participation/control rights that separate policy (on-chain) from execution (foundation operations).

Inhibitors include regulatory uncertainty around tokens/wrappers, integration costs, token concentration and governance fatigue as participation scales, and oracle dependence. These motivate governance load-balancing (delegated committees, rate-limited upgrades, time-locks and challenge windows) so decision speed and inclusiveness remain viable as the number of participants increases.

There are boundary conditions: Where tokens or specific wrappers are restricted, the HC degrades gracefully by activating onshore modules, using contractual voting and verifiable off-chain attestation registries, and limiting on-chain execution to what local rules permit, while preserving auditable governance. Data-residency constraints can be met by keeping sensitive data off-chain with hashed proofs and regulated access via local entities. Alternative neutral wrappers foundation/association/cooperative forms can be substituted provided they commit to code deference and a purpose-bound remit.

Our findings complement work arguing that decentralization disrupts platforms only when paired with robust governance design (Ladd et al., 2024). Through this lens, the HC might function as a pragmatic 'Web2.5' architecture: programmable where it helps, legally anchored where it must.

For practitioners, start with a neutral wrapper with code deference, define participation and control rights up front (director independence, election/ratification, removal for breach), adopt multi-provider oracles with override thresholds and audit trails, and plan onshore modules early for sectoral compliance. Publish token policy (allocation, vesting/lockups, for-cause clawbacks) to align contributions with benefits and limit capture.

For policymakers, recognizing code-deferent wrappers, clarifying liability/tax treatment of foundation-style entities, and offering supervisory sandboxes for oracle and token mechanisms would improve legal certainty while preserving the accountability benefits of programmable governance.

Conceptually, our contribution is twofold: (i) we articulate credible, enforceable decentralization as a socio-technical design principle that bridges code and law; and (ii) we extend hybrid organizing to multi-jurisdiction digital infrastructures via a layered architecture linking programmable governance, a neutral legal foundation, and jurisdictional modules.

## 7. Conclusion and Future Research

This paper asked: *How can a digital governance model enable trustworthy, inclusive coordination among semi-autonomous actors in complex supply chain networks?*

We proposed the Hybrid Cooperative (HC), a digitally native artifact that integrates DAO-based coordination with legal and jurisdictional modularity. Grounded in design science and organizational theory and evaluated post-case on two supply-chain traceability projects, the HC offers a viable blueprint for addressing institutional gaps where existing models (DAOs, cooperatives, platform cooperatives, foundations) struggle to scale.

The evaluation is post-case and illustrative – focused on governance in two well-documented projects as an early assessment; fuller empirical validation is left to future work.

Future research should examine (i) longitudinal, multi-site deployments to develop a process theory of HC development and use; (ii) comparative evaluations versus firm-led consortia, platform cooperatives, and DAO-native models; and (iii) experiments/simulations on quorum thresholds, delegation schemes, oracle designs, and token issuance/vesting/redemption rules. Further work should also analyze how hybrid models negotiate misalignments between on-chain governance and off-chain legal obligations, how legitimacy and trust evolve across layers, and how such systems remain adaptable under shifting regulatory, technological, and social conditions.

As IS research seeks responsible digital infrastructures, hybrid architectures like the HC offer both a conceptual lens and a practical scaffold. This work is one provisional, adaptive step toward more programmable, participatory, and policy-aware forms of organizing.

## 8. References

Axelsen, H., Jensen, J. R., & Ross, O. (2024). Do You Need a DAO? A Framework for Assessing DAO Suitability. *ECIS 2024 Proceedings*, June.

Battilana, J., Sengul, M., Pache, A. C., & Model, J. (2015). Harnessing productive tensions in hybrid organizations: The case of work integration social enterprises. *Academy of Management Journal*, 58(6), 1658–1685.

Beck, R., & Müller-Bloch, C. (2017). Blockchain as radical innovation: A framework for engaging with distributed ledgers. *Proceedings of the Annual Hawaii International Conference on System Sciences*, 2017-January, 5390–5399.

Beck, R., & Müller-Bloch, C. (2018). Governance in the Blockchain Economy: A Framework and Research Agenda. *JOURNAL OF THE ASSOCIATION FOR INFORMATION SYSTEMS*, March.

Beck, R., Schletz, M., Baggio, A., & Gentile, L. (2023). Distributed Ledger Technology for Collective Environmental Action. *Lecture Notes in Networks and Systems*, 768 LNNS, 3–15.

Brummer, C., & Seira, R. (2022). *Legal Wrappers and DAOs !* 1–31.

Carballo, R. E. (2023). Purpose-driven transformation: a holistic organization design framework for integrating societal goals into companies. *Journal of Organization Design*, 12(4), 195–215.

Davidson, S. (2023). *From CAOs to DAOs*. August.

Ebrahim, A. (2003). Accountability in practice: Mechanisms for NGOs. *World Development*, 31(5), 813–829. https://doi.org/10.1016/S0305-750X(03)00014-7

Ellinger, E. W., Gregory, R. W., Mini, T., Widjaja, T., & Henfridsson, O. (2024). Skin in the Game: the Transformational Potential of Decentralized Autonomous Organizations. *MIS Quarterly: Management Information Systems*, 48(1), 245–272.

Fairbairn, B. (1995). The Meaning of Rochdale: The Rochdale Pioneers and the Co-operative Principles. *Centre for the Study of Cooperatives Univeristy of Saskatchewan Occasional Paper Series*.

Gregor, S., & Hevner, A. (2013). *Positioning and Presenting Design Science Research for Maximum Impact*.

Gregory, R. W., Beck, R., Henfridsson, O., & Yaraghi, N. (2024). Cooperation Among Strangers: Algorithmic Enforcement of Reciprocal Exchange with Blockchain-Based Smart Contracts. *Academy of Management Review*, 00(00), 1–15.

Guo, C., & Acar, M. (2005). Understanding collaboration among nonprofit organizations: Combining resource dependency, institutional, and network perspectives. *Nonprofit and Voluntary Sector Quarterly*, 34(3), 340–361.

He, V. F., & Puranam, P. (2023). Some challenges for the "new DAOism": a comment on Klima DAO. *Journal of Organization Design*, 12(4), 293–295.

Henfridsson, O., & Bygstad, B. (2013). The Generative Mechanisms of Digital Infrastructure Evolution. *MIS Quarterly*, 37(3), 907–931.

Hevner, A. R., March, S. T., Park, J., & Ram, S. (2004). Design science in information systems research. *MIS Quarterly: Management Information Systems*, 28(1),

75–105.

ICA. (2015). Guidance Notes to the Co-operative Principles. *International Co-Operative Alliance*, 120. https://ica.coop/en/media/library/research-and-reviews/guidance-notes-cooperative-principles

Ixmeier, A., Wagner, F., & Kranz, J. (2024). Leveraging Information Systems for Environmental Sustainability and Business Value. *MIS Quarterly Executive*, *23*(1), 57–75.

Jirásek, M. (2023). Klima DAO: a crypto answer to carbon markets. *Journal of Organization Design*, *12*(4), 271–283.

Jovanovic, M., Kostić, N., Sebastian, I. M., & Sedej, T. (2022). Managing a blockchain-based platform ecosystem for industry-wide adoption: The case of TradeLens. *Technological Forecasting and Social Change*, *184*(September).

Ladd, T., Barlow, R., Giannini, B., & Pflaum, A. (2024). Decentralized Autonomous Organizations as a Threat to Centralized Platforms: Applying and Expanding Theories of Platform Competition and Disintermediation. *Proceedings of the Annual Hawaii International Conference on System Sciences*, 6340–6349.

Lumineau, F., Wang, W., & Schilke, O. (2021). Organization Science Blockchain Governance — A New Way of Organizing Collaborations ? *Organization Science*.

O'Mahony, S., & Ferraro, F. (2007). The emergence of governance in an open source community. *Academy of Management Journal*, *50*(5), 1079–1106.

Orlikowski, W. J. . (1992). The Duality of Technology : Rethinking the Concept of Technology in Organizations. *Organization Science*, *3*(3), 398–427.

Pflaum, A., Prockl, G., Bodendorf, F., & Chen, H. (2024). The Digital Supply Chain of the Future: From Drivers to Technologies and Applications. *Proceedings of the Annual Hawaii International Conference on System Sciences*, 4972–4974.

Powell, W. W. (1990). Neither Market Nor Hierarchy. *Research in Organizational Behavior*, *12*, 295–336.

Prockl, G., Roeck, D., Jensen, T., Mazumdar, S., & Mukkamala, R. R. (2022). Beyond Task-technology Fit: Exploring Network Value of Blockchain Technology Based on Two Supply Chain Cases. *Proceedings of the Annual Hawaii International Conference on System Sciences*, *2022-January*, 5040–5049.

Puranam, P., Alexy, O., & Reitzig, M. (2014). What's new about new forms of organizing? *Academy of Management Journal*, *39*(2), 162–180.

Santana, C., & Albareda, L. (2022). Blockchain and the emergence of Decentralized Autonomous Organizations (DAOs): An integrative model and research agenda. *Technological Forecasting and Social Change*, *182*(June).

Schwiderowski, J., Balle, A., & Roman, P. (2024). Crypto Tokens and Token Systems. *Information Systems Frontiers*, *26*(1), 319–332.

Shepherd, D. A., Williams, T. A., & Zhao, E. Y. (2019). A framework for exploring the degree of hybridity in entrepreneurship. *Academy of Management Perspectives*, *33*(4), 491–512.

Sternberg, H., Linan, I., Prockl, G., & Norrman, A. (2021). Tragedy of the facilitated commons: A multiple-case study of failure in systematic horizontal logistics collaboration. *Journal of Supply Chain Management*, *58*(4), 30–57.

Sternberg, H. S., Hofmann, E., & Roeck, D. (2021). The Struggle is Real: Insights from a Supply Chain Blockchain Case. *Journal of Business Logistics*, *42*(1), 71–87.

Swain, S., & Patra, M. R. (2022). A Distributed Agent-Oriented Framework for Blockchain-Enabled Supply Chain Management. *2022 IEEE International Conference on Blockchain and Distributed Systems Security, ICBDS 2022*, 1–7.

UK Law Commission. (2024). *Decentralised autonomous organisations ( DAOs ) A scoping paper*. July.

Venable, J., Pries-Heje, J., & Baskerville, R. (2016). FEDS: A Framework for Evaluation in Design Science Research. *European Journal of Information Systems*, *25*(1), 77–89.

Williamson, O. E. (1981). *The Economics of Organization : The Transaction Cost Approach*. *87*(3), 548–577.

World Bank Group. (2019). Open Source for Global Public Goods. *Open Source for Global Public Goods*.

Wright, S. A. (2021). Measuring DAO Autonomy: Lessons From Other Autonomous Systems. *IEEE Transactions on Technology and Society*, *2*(1), 43–53.

Zhu, J., & Marjanovic, O. (2020). How do platform cooperatives contribute to sustainable development goals? *26th Americas Conference on Information Systems, AMCIS 2020*, August.

Zhu, J., & Marjanovic, O. (2024). A Different Kind of Sharing Economy: A Taxonomy of Platform Cooperatives. *Proceedings of the Annual Hawaii International Conference on System Sciences*, *1*, 4174–4183.